\documentclass[11pt,a4paper]{article}
\usepackage[utf8]{inputenc}
\usepackage{amsmath,amssymb,amsthm}
\usepackage{geometry}
\usepackage{graphicx}
\usepackage{hyperref}
\usepackage{physics}
\usepackage{tikz}
\usetikzlibrary{decorations.pathmorphing,patterns,shapes,arrows,positioning}

\theoremstyle{definition}

\newcommand{\beq}[1]{\begin{equation}\label{#1}}
	\newcommand{\eeq}{\end{equation}}
\newcommand{\bear}[1]{\begin{eqnarray}\label{#1}}
	\newcommand{\ear}{\end{eqnarray}}

\newcommand{\Cinf}{ {\mathcal{C}}^\infty }
\newcommand{\Cyl}{ \mbox{\rm Cyl} }
\newcommand{\Der}{ \mbox{\rm Der}(\Cyl) }

\newcommand{\partl}{ {\partial } }

\title{Gravitation and Spacetime: Emergent from Spinor Interactions - How?}
\author{Rainer M.$^{1,2}$\\
	\small $^1$Mathematical Physics, Baku State University, Azerbaijan\\
	\small $^2$ENAMEC Institut, Würzburg, Germany\\
	\small \texttt{martin.rainer@enamec.de, martin.rainer@bsu.edu.az}}
\date{}

\begin{document}
	
	\maketitle
	
	\begin{abstract}
		Newtonian gravity arises as the nonrelativistic, static, weak-field limit of some Lorentzian spacetime geometry solving the generally covariant Einstein equations for a given matter field configuration. Spacetime geometry has a local description in the spinor basis of Penrose. The breakdown of relativistic quantum (field) theory at small distances suggests that, the Lorentzian geometry is to be modified below some regularization length. The thermodynamic correspondence, e.g. for black holes or other horizons, indicates that, Lorentzian spacetime is an emergent geometric description of an ensemble of more fundamental constituents. The independent derivations of the area law of the Bekenstein-Hawking entropy by string theory and loop quantum gravity show that, (some) properties of spacetime do not depend on the nature of its fundamental constituents (in leading order).
		
		Whether, on a fundamental scale, spacetime gravity has its own classical or quantum constituents (like e.g. in loop quantum gravity), or it is just an effective theory, deriving from expectation values of quantum matter operators (like in spinor gravity or causal fermion systems), this is still open. 
		
		We compare some very different classical and quantum approaches to spacetime geometry, all deriving in one way or another from spinors, and comment on questions for future research in order to clarify their relations.
		
		We propose that both the causal structure and the spin networks for generation of discrete geometry arise via projection from all particle spinors (fermionic and bosonic) within a causal double cone region and their spin intertwining interaction events onto a spatial section of this double cone region.   
		
	\end{abstract}
	
	\section{Introduction}
	
	General relativity describes gravitation as the curvature of a four-dimensional Lorentzian manifold $(M, g_{\mu\nu})$, where the Einstein field equations
	\begin{equation}
		R_{\mu\nu} - \frac{1}{2}R g_{\mu\nu}  = 8\pi G T_{\mu\nu}
		\label{eq:einstein}
	\end{equation}
	relate the pure (without cosmological constant) geometry (left-hand side)  to matter (outside vacuum) content (right-hand side).  Quantum mechanics and general relativity become definitely in conflict at length scales as short as the Planck scale, when quantum uncertainty resolves the classical spacetime structure.
	
	Evidence that spacetime geometry might be emergent rather than fundamental comes from the thermodynamic nature of black hole geometry,
	in particular the black hole entropy \cite{bekenstein1973,hawking1975}
	\begin{equation}
		S_{BH} = \frac{k_B A}{4 l_P^2}
		\label{eq:bh_entropy}
	\end{equation}
	where $A$ is the horizon area.  The area law \eqref{eq:bh_entropy} shows that, the information content of a bounded region (e.g. a horizon) relates to the boundary area rather than its volume. This hints at a fundamental holographic principle \cite{thooft1993,susskind1995} applying to black holes and more general horizons.  
	
	Furthermore, a purely classical derivation of gravity as an entropic force \cite{verlinde2011}, suggests that, spacetime geometry may indeed be an emergent phenomenon. This view is further supported by considerations of entanglement entropy in quantum field theory, and the derivation of obtain Einstein's equations from thermodynamic relations.
	
	The thermodynamic nature of gravity and general relativity suggests that spacetime geometry encodes macroscopic ensemble information about more fundamental microscopic degrees of freedom. Although applying very different basic structures, namely strings and holonomy loops,
	both string theory \cite{strominger1996} and loop quantum gravity \cite{rovelli1996} succeeded to derive the area law \eqref{eq:bh_entropy} independently. Therefore some aspects of spacetime geometry may be independent of the detailed nature of the fundamental microscopic constituents, which ultimately generate gravity and spacetime. 
	
	Therefore, the question arises: what are the fundamental constituents of gravity at the very small scale? Particularly interesting candidates are spinors. Mathematically they are more fundamental than tensors and can describe locally the classical spacetime geometry \cite{penrose1984}. The Lorentzian metric, and any other tensor, can be represented as a composite object from spinors. 
	This spinor formalism essentially exploits the double cover relationship between $SL(2,\mathbb{C})$ and the Lorentz group $SO^+(3,1)$.
	
	The viewpoint that, spinors are the ultimate constituents of spacetime is supported on the matter side by the fact that, all known elementary fields of the standard model belong to a spinor representation of spin $0$, $\frac{1}{2}$ or $1$. While for the bosonic matter fields a tensor representation is available, the  fermionic matter fields (quarks, leptons) can only be represented 
	by spin $\frac{1}{2}$ spinors. Since the matter side is essentially spinorial, a spinorial nature of spacetime geometry might be natural, too.    
	
	Indeed, already classically,  local spacetime geometry can be cast in terms of pairs of spinors\cite{penrose1984}, where locally each vector and covector from a tensor representation can be soldered to spinor pair from two representations conjugate to each other.   
	Ashtekar\cite{ashtekar1986,ashtekar1987} then obtained a further reformulation of general relativity via $SU(2)$ connections and spinorial variables.
	
	Loop Quantum Gravity (LQG) 
	is based on $SU(2)$-holonomies on graphs embedded in a $3$-dimensional spatial hypersurface of a globally hyperbolic manifold. The edges of the graph intersect $2$-dimensional surfaces of transversal flux. Canonical quantization of area and volume yields geometric operators, which in a holonomy loop representation on the graph  have discrete spectra in Planck units \cite{rovelli2004,thiemann2007}. Space(time) geometry emerges from quantum spin networks, where $SU(2)$-representations of spin $j_e$ live on edges $e$ of the graph. The vertices of the graph represent the interaction of their adjacent spinor representations, which have to satisfy Clebsch-Gordon-like sum rules realized by intertwiners of the adjacent
	spin $j_e$ representations.  
	
	Unlike LQG, the very different {Causal Fermion Systems} (CFS) approach \cite{finster2016,finster2020} proposes a classical emergent spacetime without quantization of gravity at small distances. Rather spacetime points are assumed to emerge from a correlation structure given by fermionic projector states, which derive from a Hilbert space of solutions of the Dirac equation. A particular regularized (i.e. non lightlike) notion of causality is a priori built in by the spectral properties of the projectors.
	A remarkable feature of this approach is that (in leading orders of the regulator) it may recover QFT on a classical curved spacetime, if a minimum of $3$ fermion generations is taken into account. We will however not follow this approach here, but just keep in mind that, it supports the view that fermionic spinors play a fundamental role as ingredients from which classical spacetime may emerge.
	
	This paper proceeds from the classical local spinor representations of spacetime, via $3+1$-decomposition to the holonomy loop representation and the spin network representation of loop quantum gravity.
	A metric-independent concept of causal structure is introduced with its relevance for local causal quantum theory. In that context, the spin network structure is examined for its relation to space(time) geometry and particle content.
	Finally the ontological question of the ultimate constituents of spacetime is discussed.
	
	\section{Classical Spacetime Geometry in Tetrad and Spinor Variables}
	
	\subsection{Tetrad Frame Fields}
	With tetrad frame fields $e^a_\mu(x)$ soldering locally any spacetime tangent vector $v^\mu$ to a vector $v^a$ in Minkowski space,
	the spacetime metric may locally around any point $x$ be recovered as
	\begin{equation}
		g_{\mu\nu}(x) = e^a_\mu(x) e^b_\nu(x) \eta_{ab} .
		\label{eq:metric_tetrad}
	\end{equation}
	
	\subsection{Spinor Frame Fields}
	The Lorentz group $SO^+(3,1)$ has a double cover $SL(2,\mathbb{C})$, the group of $2\times 2$ complex matrices with determinant one. A two-component spinor $\xi^A$ ($A = 0,1$) transforms as
	\begin{equation}
		\xi^A \to \Lambda^A{}_B \xi^B, \quad \Lambda \in SL(2,\mathbb{C})
		\label{eq:spinor_transf}
	\end{equation}
	The fundamental antisymmetric tensor $\epsilon_{AB}$ with $\epsilon_{01} = 1$ allows index raising and lowering:
	\begin{equation}
		\xi_A = \epsilon_{AB}\xi^B, \quad \xi^A = \epsilon^{AB}\xi_B
	\end{equation}	
	Complex conjugate spinors are denoted with primed indices $\overline{\xi^A} = \xi^{A'}$. The Infeld-van der Waerden symbol $\sigma^a_{AA'}$ establish a correspondence between a Minkowski vector and a pair of conjugate spinors \cite{penrose1984}:
	\begin{equation}
		v^a =  \sigma^a_{AA'} v^{AA'} 
		\label{eq:vector_spinor}
	\end{equation}
	with some suitable bispinor $v^{AA'}:=\xi^A\eta^{A'}$. With its dual $\sigma^{AA'}_a$,
	the Minkowski metric tensor reads
	\begin{equation}
		\eta_{ab} = \sigma^{AA'}_a\sigma^{BB'}_b\epsilon_{AB}\epsilon_{A'B'} .
	\end{equation}
	Similarly as in (\ref{eq:vector_spinor}), 
	locally around any spacetime point $x$ a bispinorial tetrad field $\sigma^{AA'}_\mu(x)$ yields 
	the metric (\ref{eq:metric_tetrad}) as 
	\begin{equation}
		g_{\mu\nu}(x) = \sigma^{AA'}_\mu(x) \sigma^{BB'}_\nu(x) \epsilon_{AB}\epsilon_{A'B'} .
		\label{eq:metric_spinorial}
	\end{equation}

	\subsection{Curvature Spinors}
	All curvature tensors can be expressed by substituting each tensorial vector or form component by an appropriate bispinorial one.
	E.g. the Riemann curvature decomposes into Weyl conformal curvature and Ricci curvature. In spinor form, the Weyl tensor corresponds to a totally symmetric spinor $\Psi_{ABCD}$ \cite{penrose1984}:
	\begin{equation}
		C_{\mu\nu\rho\sigma} = \Psi_{ABCD} \epsilon_{A'B'} \epsilon_{C'D'} \sigma^{AA'}_\mu \sigma^{BB'}_\nu \sigma^{CC'}_\rho \sigma^{DD'}_\sigma + \text{c.c.}
		\label{eq:weyl_spinor}
	\end{equation}
	Physically this demonstrates that, causal structure arises essentially as a spinorial structure.
	
	Likewise, the trace-free Ricci tensor is encoded in a trace-free Ricci spinor $\Phi_{ABA'B'}$ and the scalar curvature $R$ arises as a spinorial contraction. Physically this relates to the fact, that ultimately all matter is spinorial, fermionic one with spin $\frac{1}{2}$ bosonic one with spin $1$ for the gauge bosons and  spin $0$ for the Higgs.  
	
	\subsection{3+1 Decomposition of Spacetime in Spinor Variables}
	
	To formulate gravity canonically, spacetime must be foliated into spacelike hypersurfaces $\Sigma_t$ \cite{arnowitt2008}. The metric takes the ADM form:
	\begin{equation}
		ds^2 = -N^2 dt^2 + h_{ij}(dx^i + N^i dt)(dx^j + N^j dt)
		\label{eq:adm_metric}
	\end{equation}
	where $N$ is the lapse function, $N^i$ the shift vector, and $h_{ij}$ the induced spatial metric.
	With $D_i$ denoting the covariant derivative compatible with $h_{ij}$,
	the extrinsic curvature of $\Sigma_t$ embedded in spacetime is
	\begin{equation}
		K_{ij} = \frac{1}{2N}(\partial_t h_{ij} - D_i N_j - D_j N_i) ,
		\label{eq:extrinsic_k}
	\end{equation}
	and the momentum conjugate to $h_{ij}$ is
	\begin{equation}
		\pi^{ij} = \sqrt{h}(K^{ij} - h^{ij}K) .
		\label{eq:adm_momentum}
	\end{equation}
	The phase space variables $(h_{ij}, \pi^{ij})$ are known as
	ADM canonical variables. 
	The ADM Hamiltonian in terms of these variables is  
	\begin{equation}
		H_{ADM} = \int_\Sigma d^3x (N\mathcal{H} + N^i\mathcal{H}_i) .
	\end{equation}
	If the Hamiltonian and $3$-diffeomorphism constraints
	\begin{align}
		\mathcal{H} &= \frac{1}{\sqrt{h}}(\pi^{ij}\pi_{ij} - \frac{1}{2}\pi^2) - \sqrt{h}R^{(3)} = 0 \label{eq:ham_constraint}\\
		\mathcal{H}_i &= -2D_j\pi^j_i = 0 \label{eq:diff_constraint}
	\end{align}
	are satisfied, arbitrariness of lapse $N$ and shift $N^i$ is enabling general covariance in spite of the $3+1$-decomposition. 
	
	Let $n^\mu$ be the future-pointing unit normal to $\Sigma_t$:
	\begin{equation}
		n^\mu n_\mu = -1
	\end{equation}
	The projection tensor onto $\Sigma_t$ is
	\begin{equation}
		h_{\mu\nu} = g_{\mu\nu} + n_\mu n_\nu
	\end{equation}
	In spinor notation, the unit normal becomes $n^{AA'} = t^A \bar{t}^{A'}$, where $t^A$ is a normalized spinor field. 
	The spatial triad $e^i_a$ on $\Sigma_t$ (with internal $SO(3)$ index $a$) can then be lifted to $SU(2)$ spinors. 
	The additional $SU(2)$ structure on the spatial slice requires an additional gauge constraint. 
	
	\section{Classical Spacetime Geometry in Ashtekar Variables}
	Ashtekar introduced a reformulation using the self-dual part of the spin connection \cite{ashtekar1986,ashtekar1987}. 
	With a spatial triad $e^i_a$ ($a$ is now an internal $SU(2)$ index) on hypersurface $\Sigma_t$, satisfying
	\begin{equation}
		h_{ij} = e^a_i e^b_j \delta_{ab} ,
	\end{equation}
	the densitized triad (momentum variable) is
	\begin{equation}
		E^a_i = \sqrt{h} e^a_i
		\label{eq:densitized_triad}
	\end{equation}
	The self-dual Ashtekar connection (configuration variable) is
	\begin{equation}
		A^a_i = \Gamma^a_i - i K^a_i  ,
		\label{eq:ashtekar_connection}
	\end{equation}
	where $\Gamma^a_i$ is the spin connection compatible with $e^a_i$, and $K^a_i = K_{ij}e^{aj}$ is the extrinsic curvature in triad form. The factor of $i$ selects the self-dual part.
	
	The canonical Poisson brackets are
	\begin{equation}
		\{A^a_i(x), E^j_b(y)\} = \delta^a_b\delta^j_i\delta^3(x,y)
		\label{eq:poisson_ashtekar}
	\end{equation}
	The Einstein constraints become remarkably simple, if the connection is chosen to be the selfdual one (\ref{eq:ashtekar_connection}). 
	
	{Gauss constraint (gauge invariance):}
	\begin{equation}
		\mathcal{G}_a = D_i E^i_a = \partial_i E^i_a + \epsilon_{ab}^c A^b_i E^i_c = 0
		\label{eq:gauss}
	\end{equation}
	
	{Diffeomorphism constraint:}
	\begin{equation}
		\mathcal{H}_i = F^a_{ij}E^a_j = 0
		\label{eq:diffeo}
	\end{equation}
	where $F^a_{ij} = \partial_i A^a_j - \partial_j A^a_i + \epsilon^{abc}A^b_i A^c_j$ is the field strength.
	
	{Scalar (Hamiltonian) constraint:}
	\begin{equation}
		\mathcal{H} = \frac{\epsilon_{abc}E^a_i E^b_j}{\sqrt{\det E}}F^c_{ij} = 0
		\label{eq:hamiltonian_lqg}
	\end{equation}
	All constraints are polynomial in the fundamental variables. In particular (\ref{eq:hamiltonian_lqg}) becomes more complicated, if a different e.g. real connection was chosen. 
	The selfdual connection $A^a_i$ is complex: 
	\begin{equation}
		\text{Im}(A^a_i) = -K^a_i
	\end{equation}
	For real general relativity then, reality conditions must be imposed. Various strategies address this, including a real connection with the Immirzi parameter $\gamma$ \cite{barbero1995} and the real Lorentzian formulation \cite{thiemann2007}.

	\section{Holonomy and Flux Variables}
	
	
	In gauge theory, parallel transport along a path $\gamma$ is described by the holonomy:
	\begin{equation}
		h_\gamma[A] = \mathcal{P}\exp\left(\int_\gamma A^a_i \tau_a dx^i\right) \in SU(2)
		\label{eq:holonomy}
	\end{equation}
	where $\tau_a = -i\sigma_a/2$ are $SU(2)$ generators and $\mathcal{P}$ denotes path-ordering.
	
	For quantum gravity, holonomies are more fundamental than connections. They are gauge-invariant (under closed loops), well-defined as operators, and avoid ultraviolet divergences since they "smear" the connection over finite paths \cite{rovelli2004}.
	
	
	A conjugate variable to the holonomy is the flux through a surface $S$, intersecting $\gamma$ transversally,
	\begin{equation}
		E_S^a[E] = \int_S \epsilon^{ijk} E^a_i dS_{jk} .
		\label{eq:flux}
	\end{equation}
	
	The Poisson bracket between holonomy and flux is 
	\begin{equation}
		\{h_\gamma[A], E_S^a[E]\} = \sum_{p \in \gamma \cap S} \tau^a h_{\gamma_p}[A]
		\label{eq:holonomy_flux_poisson}
	\end{equation}
	where the sum is over intersection points $p$ of the path $\gamma$ with surface $S$ at its segment $\gamma_p$, with	 orientations taken into account.
	
	\section{Spin Network Representation}
	
	\subsection{Definition of Spin Networks}
	
	A spin network $\Gamma = (\gamma, j_e, i_v)$ consists of \cite{penrose1971,rovelli1995}:
	\begin{itemize}
		\item An oriented graph $\gamma$ embedded in $\Sigma$, with edges $e$ and vertices $v$
		\item Spins $j_e \in \{0, 1/2, 1, 3/2, \ldots\}$ labeling each edge
		\item Intertwiners $i_v$ at each vertex (i.e. invariant recouplings of the spin representations 
		$R^{(j_e)}$ of the oriented incoming and outgoing edges at each vertex)
	\end{itemize}
	
	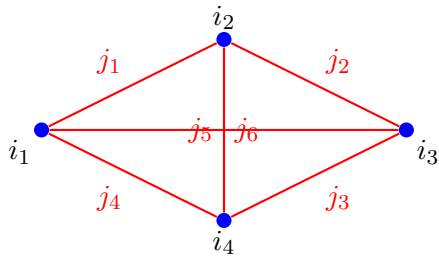
\begin{figure}[h]
		\centering
		\begin{tikzpicture}[scale=1.2]
			\node[circle,fill=blue,inner sep=2pt] (v1) at (0,0) {};
			\node[circle,fill=blue,inner sep=2pt] (v2) at (2,1) {};
			\node[circle,fill=blue,inner sep=2pt] (v3) at (4,0) {};
			\node[circle,fill=blue,inner sep=2pt] (v4) at (2,-1) {};
			
			\draw[thick,red] (v1) -- node[above left] {$j_1$} (v2);
			\draw[thick,red] (v2) -- node[above right] {$j_2$} (v3);
			\draw[thick,red] (v3) -- node[below right] {$j_3$} (v4);
			\draw[thick,red] (v4) -- node[below left] {$j_4$} (v1);
			\draw[thick,red] (v1) -- node[left] {$j_5$} (v3);
			\draw[thick,red] (v2) -- node[right] {$j_6$} (v4);
			
			\node[below left] at (v1) {$i_1$};
			\node[above] at (v2) {$i_2$};
			\node[below right] at (v3) {$i_3$};
			\node[below] at (v4) {$i_4$};
		\end{tikzpicture}
		\caption{Example of a spin network. Vertices (blue dots) are labeled with intertwiners $i_v$, and edges (red lines) carry spins $j_e \in \mathbb{N}/2$.}
		\label{fig:spin_network}
	\end{figure}
	
	A spin network state can be constructed on the spin-labeled graph as 
	\begin{equation}
		\Psi_\Gamma[A] = \prod_{v \in \gamma} i_v \left[ \prod_{e \in v} R^{(j_e)} (h_e[A]) \right]
		\label{eq:spin_network_state}
	\end{equation}
	$\Psi_\Gamma[A]$ is invariant under diffeomorphisms of the graph $\gamma$
	and gauge invariant under all spin $j_e$ gauge groups involved.
	
	\subsection{Quantum Geometry of Spin Networks}
	LQG takes the point of view, that spacetime geometry is to be quantized canonically.
	
	The  kinematical Hilbert space for quantum geometry may be generated from the spin network states (\ref{eq:spin_network_state}).
	Upon quantization, holonomies and fluxes become operators thereon \cite{ashtekar2004}:
	\begin{align}
		\hat{h}_\gamma[A] \Psi[A] &= h_\gamma[A] \Psi[A] \label{eq:holonomy_op}\\
		\hat{E}_S^a[E] \Psi[A] &= -i \frac{\delta}{\delta A_S^a} \Psi[A] \label{eq:flux_op} .
	\end{align}
	Remarkably, geometric operators have discrete spectra when acting on spin network states \cite{rovelli1995,ashtekar1997}. E.g. for a surface $S$, the area is:
	\begin{equation}
		\hat{A}(S) \Psi_\Gamma = 8\pi \gamma l_c^2 \sum_{p \in S \cap \Gamma} \sqrt{j_p(j_p+1)} \, \Psi_\Gamma
		\label{eq:area_operator}
	\end{equation}
	where the common standard choice for the critical length scale is $l_c =l_P = \sqrt{G\hbar/c^3}$, the Planck length, and $\gamma$ is the Immirzi parameter. As pointed out in \cite{{rainer1999length}}, the value of $\gamma$ may be set according to the  relevant scale $l_c>l_P$  for the resulting discrete spatial geometry. In the following section, we propose that, the density of particle interactions is setting the scale $l_c$.
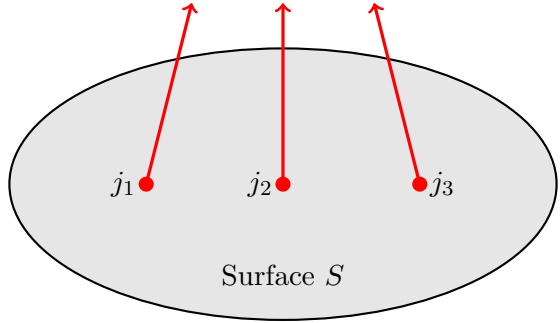
	\begin{figure}[h]
		\centering
		\begin{tikzpicture}[scale=1.2]
			\draw[thick,fill=gray!20] (0,0) ellipse (3cm and 1.5cm);
			\node at (0,-1.0) {Surface $S$};
			
			\draw[very thick,red,->] 
			(-1.5,0) -- (-1,2);
			\draw[very thick,red,->] 
			(0,0) -- (0,2);
			\draw[very thick,red,->] 
			(1.5,0) -- (1,2);
			
			\node[circle,fill=red,inner sep=2pt] (p1) at (-1.5,0) {};
			\node[circle,fill=red,inner sep=2pt] (p2) at (0,0) {};
			\node[circle,fill=red,inner sep=2pt] (p3) at (1.5,0) {};
			
			\node[left] at (-1.5,0) {$j_1$};
			\node[left] at (0,0) {$j_2$};
			\node[right] at (1.5,0) {$j_3$};
			
			\node at (0,-2.5) {$A(S) = 8\pi\gamma l_P^2 \sum_p \sqrt{j_p(j_p+1)}$};
		\end{tikzpicture}
		\caption{Spin network edges piercing a surface $S$. Each puncture contributes to the quantized area according to its spin label.}
		\label{fig:surface_puncture}
	\end{figure}

	\subsection{Surface Entropy from Spin Networks}
	
	The black hole entropy can be derived by counting spin network states that pierce the horizon with the correct area \cite{rovelli1996,ashtekar1997}. For large area $A \gg l_P^2$:
	\begin{equation}
		S = \frac{k_B}{\hbar} \ln \Omega(A) \approx \frac{k_B c^3 A}{4G\hbar}
		\label{eq:lqg_entropy}
	\end{equation}
	matching Equation \eqref{eq:bh_entropy}, with corrections depending on $\gamma$ and spin statistics.
	The microstate counting function $\Omega(A)$ counts the number of distinct spin network configurations that yield a horizon area eigenvalue equal to $A$. Since the area operator has discrete spectrum (Equation \eqref{eq:area_operator}), only specific spin network configurations with punctures $\{j_p\}$ satisfying
	\begin{equation}
		A = 8\pi \gamma l_P^2 \sum_{p} \sqrt{j_p(j_p+1)}
	\end{equation}
	contribute to $\Omega(A)$ with spin values $j_p$ at puncture $p$.
	Puncturing edges with $j_p=\frac{1}{2}$ orrespond to fermions.
	Hypothetical spin $0$ punctures do not yield any area. 
	A single spin $\frac{1}{2}$  puncture yields the minimal area gap 
	$A_{\text{gap}}=4\sqrt{3}\pi \gamma l_P^2 $.   
	
	In the canonical LQG approach the Immirzi parameter $\gamma$, mixing connection and extrinsic curvature into the canonical connection, is a real number. As discussed in \cite{rainer1999length}, the geometric operators depend on a fundamental length scale $\ell$ related to the Immirzi parameter via $\ell = \sqrt{\gamma} l_P$. Therefore, with modified $\gamma_c$, the relevant critical length $l_c$ can be much larger than $l_P$, but nevertheless
	still match the correct black hole entropy with
	\begin{equation}
		A = 8\pi \gamma_c l_c^2 \sum_{p} \sqrt{j_p(j_p+1)} .
		\label{eq:area_discrete}
	\end{equation}
	A given minimal area quantum
	$A_{\text{gap}}$
	fixes the discrete structure of the surface states. The number of quanta given by the ratio $A/A_{\text{gap}}$ then yields the entropy counting \cite{rainer1999length}.
	
	Beyond the leading area-proportional term, loop quantum gravity also predicts subleading logarithmic corrections to black hole entropy~\cite{meissner2004}:
	\begin{equation}
		\frac{S}{k_B} = \frac{1}{4}\frac{A}{ l_P^2} - \frac{1}{2} \ln\left(\frac{A}{l_P^2}\right) + O(1)
	\end{equation}
	These quantum corrections are universal, independent of the detailed black hole parameters.
	These corrections are a prediction of LQG, which is experimentally testable in principle. 
	
	\section{Causal Structure and Double Cones}
	The Haag-Kastler axioms of algebraic quantum (field) theory on Minkowski space are formulated on a net of local algebras of observables \cite{haag1992}. The local topological set structure underlying the algebra respects relativistic causality, if the topological basis is chosen as nested compact double cones or their nested open interiors. 
	The topological and causal structure on a differentiable manifold can be defined via cone structures~\cite{rainer1999cones}. On a differentiable manifold admitting such a cone structure, causally complete double cone regions are the natural domains for local observables. In this way, the Haag-Kastler setting can be extended to a diffeomorphism-invariant causal setting for quantum theory~\cite{rainer2000haag}.
	
	\subsection{Compact double cones}
	
	\begin{figure}[h]
		\centering
		\begin{tikzpicture}[scale=2.5]
			
			\draw[thick] (0,-1.5) -- (0,1.5);
			
			
			\draw[thick] (0,0) -- (1.5,0);
			
			
			\draw[thick] (0,1.5) -- (1.5,0);
			
			\draw[thick] (0,-1.5) -- (1.5,0);
			
			\filldraw (0,1.5) circle (1.5pt);   
			\filldraw (0,-1.5) circle (1.5pt);  
			\filldraw (1.5,0) circle (1.5pt);   
			
			\node[left] at (-0.05,1.6) {$i^+(\rho)$};
			
			\node[right] at (0.8,0.7) {$\mathcal{J}^+(\rho)$};
			
			\node[above right] at (1.55,0.05) {$i^0(\rho) \cong S^2$};
			
			\node[right] at (0.8,-0.7) {$\mathcal{J}^-(\rho)$};
			
			\node[left] at (-0.05,-1.6) {$i^-(\rho)$};
			
			\node[above] at (0.7,0.0) {$\Sigma_0(\rho)$};
			
			\node[left] at (0.0,0.0) {$O$};

			
			
		\end{tikzpicture}
		\caption{
			The double cone $\mathcal{D}(i^-(\rho),i^+(\rho))$, with future timelike $i^+(\rho)$, past timelike $i^-(\rho)$, and spatial $i^0(\rho) \cong S^2$, future light cone $\mathcal{J}^-(\rho)$ of $i^-(\rho)$ 
			and past light cone $\mathcal{J}^+(\rho)$ of $i^+(\rho)$,  
			depicted for a certain value of $\rho>0$. $\Sigma_0(\rho)$ denotes the spatial section containing the origin $O$, the causal point of reference.
		}
		\label{fig:doublecone}
	\end{figure}
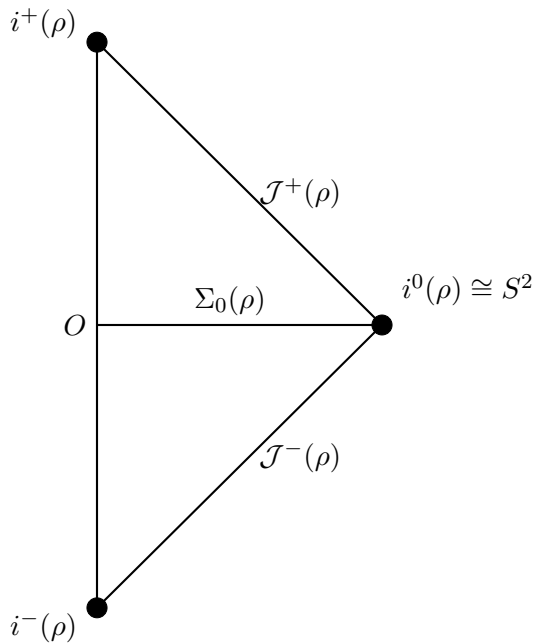
	
	A (compact) double cone (also called causal diamond) $\mathcal{D}(p,q)$ is the intersection of the future light cone of event $p$ and the past light cone of event $q$:
	\begin{equation}
		\mathcal{D}(p,q) = \mathcal{J}^+(p) \cap \mathcal{J}^-(q) .
		\label{eq:double_cone}
	\end{equation}
	$\mathcal{D}(p,q)$ contains the maximal domain for observable interaction, which can be both, influenced by $q$ and detected by $p$. Due to causal construction, events outside a double cone cannot influence neither particles nor geometry within it.

	A causal topological basis of double cones $\mathcal{D}(\rho):=\mathcal{D}(i^-(\rho),i^+(\rho))$  can labeled by a parameter $\rho>0$, which in Minkowski space becomes a radial variable (see Fig. \ref{fig:doublecone}).
	(Alternatively, around a reference point $0$, a time-like parameter $\tau>0$, with $(\pm\tau,0)=i^\pm(\rho(\tau))$, might label the topological basis of double cones as
	$\mathcal{D}(\tau):=\mathcal{D}((-\tau,0),(\tau,0))$.) In the following, when a single double cone is considered, we omit its label for simplicity of notation.

	\subsection{Spin Networks and Particle Spinors within Causal Double Cones}
	For a given $3+1$ decomposition, we consider a spin network $\Gamma$ for a graph $\gamma\subset\Sigma_0\subset \mathcal{D}$ of a spin network inside a spatial  hypersurface $\Sigma_0$ of a compact double cone $\mathcal{D}$. 
	This framework allows loop quantum gravity to be formulated in a diffeomorphism-invariant algebraic QFT setting, with spin network states accommodating isotony and causality~\cite{rainer2000lqg}.
	
	Although spin networks recently were considered mainly for construction of quantum geometry, we want to consider them also
	from a particle physics perspective. 
	
	The labeled graph $\Gamma$ may be considered as an entanglement diagram for spinors of representation $j_e$
	on its edges, and intertwiners $i_v$  at vertices $v$.
	This entanglement diagram can be seen as a  projection $\mathcal{D}\rightarrow \Sigma_0$ of the path network of all spinors and their interactions in $\mathcal{D}$, where spinor paths project to edges $e$ and their interaction events project to vertices $v$. 
	
	In this picture, a spinor of representation $j_e$ traces a future-directed path in $\mathcal{D}$, which projects to an oriented edge $e$
	within $\Gamma$. An interaction event of $n$ spinors at a point $p\in \mathcal{D}$ projects to an intertwiner vertex $v$ of $\Gamma$. Each intertwiner $i_v$ at a $v$
	encodes the entanglement of the  $n$ spinors.
	
	The spin representations of the standard model (SM) are the following: $j=0$ scalar (Higgs), $j=\frac{1}{2}$ chiral fermion (lepton or quark), $j=1$ bosons (photon, weak bosons, gluons).
	Note that, bosons may also be represented as spinors, e.g. a gauge boson may be written as $j=1$ generalization of a pair of chiral Weyl spinors, or as a generalized Dirac spinor. However, a massless boson, e.g. the electromagnetic field, has only $2$ helicity degrees of freedom rather than all spinor degrees of freedom. Hence the spinor representation of a massless boson is restricted to $m=\pm j$.  
	
	Given this additional interpretation of a spin network, the spinor paths of the standard model particles within $\mathcal{D}$ might at the same time also project to a spin network for the geometry of their space(time) they live in. Even more, space(time) may not need separate spinorial constituents, because the content of particles and their interaction events may be enough to generate the geometry of their space(time).
	
	Scalar particle fields of spin $0$ do not contribute to (area) geometry. The spin $0$ field of the SM is the Higgs field. It is supposed to attribute the mass for all massive SM particles, in particular the lepo    ns and quarks, but also the bosons of the weak interaction. 
	The Higgs mass is  fixed by a Higgs field vacuum expectation
	\begin{align}
		m_{\text{Higgs}}^2 &= \lambda\langle H^\dagger H\rangle .
	\end{align}
	We proposed the Higgs field $H$  to be generated discretely by spin $0$ edges of the spin network. They do not contribute own patches to (area) geometry, but rather make up the Higgs field, which then provides a Compton clock and the masses of the particles. In this way, spin $0$ edges set the geometric scale $l_c$ fixing the size of the area patches in (\ref{eq:area_discrete}) induced by edges with spin $j>0$.  
	
	At least the spin $1$ photon is a massless SM particle. The path of such a massless particle is per definition a geodesic of the yet unknown spacetime geometry. The more such gedesic paths are known, the more the causal (cone) structure of the space(time) geometry is already pinned down in $M$. 
	
	A massive particle provides a Compton clock inside the double cone $\mathcal{D}(p,q)$ along its timelike path between the event $p$, where it is created, and the event $q$, where it is deleted. 
	The massive particles in $\mathcal{D}(p,q)$ then fix the time and length scales of the geometry, breaking its conformal invariance. 
	
	In this way, the particles of the standard model play together to form a classical spacetime geometry,
	which can accomodate them. For this purpose, it not required to quantize geometry in the canonical 
	sense of LQG. However, the spin networks, which are underlying in particular the supposed quantum states of LQG, here appear as a natural UV-regulator of the classical geometry. Moreover, the spin networks need not to be introduced as an extra structure for small scale space(time), but they can be obtained naturally form the physical net of interaction events between the SM model spinor paths,
	keeping in mind, that both bosons and fermions have a spinor representation.

	%
	%
	%
	%
	%
	
	\subsection{Conformal Geometry and Spin Networks}
	
	Let us consider conformal transformation of geometry given by
	\begin{equation}
		g^{(c)}_{\mu \nu} := \phi^2 g_{\mu \nu} .
	\end{equation}
	The conformal (Holst) action is 
	\begin{equation}
		S_H^{(c)} = \int \frac{1}{2}e^{(c)} e_a^{\mu(c)}e_b^{\nu(c)}\left(\Omega_{\mu \nu}^{ab(c)} + \frac{1}{\gamma}{}^*\Omega_{\mu \nu}^{ab(c)}\right)
	\end{equation}
	where $\gamma$ is the Immirzi parameter,
	Latin indices $a,b,\ldots$ label the representation basis for the gauge group $SL(2,\mathbb{C})$, and $\Omega_{\mu \nu}^{ab(c)}$ is the curvature of the $SL(2,\mathbb{C})$ connection $\omega_\mu^{ab(c)}$.
	We follow here the formulation as in \cite{campiglia2016conformalLQG}.
	After $3+1$ decomposition, we have triads and connections transforming conformally as
	\begin{align}
		e_i^{a(c)} &= \phi e_i^a \\
		e^{i(c)}_a &= \frac{e^i_a}{\phi} \\
		\omega_i^{ab(c)} &= \omega_i^{ab} - 2\phi^{-1}\partial_j\phi e^{j[a} e^{b]}_i
	\end{align}
	with spatial Latin indices $i,j,\ldots$. The Poisson  bracket of conformal extrinsic curvature and conformally invariant densitized triad is
	\begin{equation}
		\{K_i^{a(c)}(x), E_b^{j(c)}(y)\} = \delta_i^j\delta^a_b\delta^3(x,y) .
	\end{equation}
	The dilaton field and its conjugate momentum are an additional pair with
	\begin{align}
		\{\phi(x), \pi^{(c)}(y)\} &= \delta^3(x,y) ,\\
		\pi^{(c)} &\approx 0 .
	\end{align}
	Conformal invariance of the action implies the vanishing of the latter.
	
	Transformation back to original geometrical variables
	\begin{align}
		E_a^i &= \phi^{-2}E_a^{i(c)} \\
		K^a_i &= \kappa\phi^2 K_i^{a(c)} \\
		\pi &= \pi^{(c)} - 2\phi^{-1}K_i^{a(c)}E_a^{i(c)}
	\end{align}
	gives the geometric Poisson brackets
	\begin{equation}
		\{K_i^a(x), E_b^j\} = \kappa\delta_a^b\delta^j_i\delta^3(x,y), \quad \{\phi(x),\pi(y)\} = \delta^3(x,y)
	\end{equation}
	and a conformal constraint quantity
	\begin{equation}
		\phi\pi^{(c)} = + \pi\phi+\frac{2}{\kappa}K_i^a E_a^i 
	\end{equation}
	With real connection variables $A_i^a = \gamma K_i^a + \Gamma_i^a$ the
	scalar (Hamiltonian) and diffeomorphism constraints are conformally modified to
	\bear
	& H= & \frac{\phi^2}{2}\epsilon_a^{bc}E_l^i E_c^j\left[F_{ij}^a - \left(\gamma^2 +1) (\kappa\phi^2)^{-2}\right)K_i^b K_j^c\epsilon_{bc}^a\right] \nonumber\\
	& + E^{ia}E_a^j [- \partial_i\phi\partial_j\phi + 2(\nabla_i\partial_j\phi)\phi] ,
	\ear
	\begin{equation}
		\mathcal{H}_i = \frac{1}{\gamma}F_{ij}^a E_a^j + \pi\partial_i\phi .
	\end{equation}
	The conformal Ashtekar connection 
	\begin{equation}
		A_i^{a(c)} = \Gamma_i^{a(c)} + \gamma K_i^{a(c)} \quad, \quad A_i^{(c)} =A_i^{a(c)} \tau_a  
	\end{equation}
	defines a conformal holonomy along an oriented path $\eta$ as
	\begin{equation}
		h_\eta[A^{(c)}]=U(A^{(c)},\eta) = P\exp\int_\eta A_i^{(c)}dx^i \in SU(2).
	\end{equation}
	Under $SU(2)$ gauge transformations, conformal holonomies
	transform as usual via
	\begin{equation} 
		U(A^{(c)},\eta) \to \Lambda(x_f^\eta)U(A^{(c)},\eta)\Lambda^{-1}(x_i^\eta) ,
	\end{equation}
	where $x_f^\eta$ and $x_i^\eta$ are the end and starting points of the path $\eta$.
	
	The conformal holonomies yield cylindrical conformal functions related 
	to $\Gamma = (\gamma, j_e, i_v)$ as
	\begin{equation}
		\Psi_\Gamma[A^{(c)}] = \prod_{v \in \gamma} i_v \left[ \prod_{e \in v} R^{(j_e)} (h_e[A^{(c)}]) \right] ,
	\end{equation}
	where $R^{(j)}$ is the spin $j$ representation of the gauge group $SU(2)$ and $i_v$ is the $SU(2)$-intertwiner of the representations adjacent to vertex $v\in\gamma$. (Here we neglect the gauge group $G_{SM}$ of the SM. In a full approach the relevant $G_{SM}$ has to be added consistently on top. For the spin network considered here, bosonic and fermionic spinors are considered equally.)
	The resulting conformal  spin network states live on a scale free topological graph $\gamma$ within a spatial projection section  $\Sigma_0\subset\mathcal{D}$ within a compact causal doublecone $\mathcal{D}$. The edges of the graph represent the particle spinors and the vertices their interaction  entanglement.
	
	Gauge fixing $\phi(x) = \phi_0 $ by (the Compton clock of) the (massive) Higgs field, breaks the conformal invariance and fixes the geometric scale $l_c$, in particular for the area (\ref{eq:area_discrete}).  
	The Newton constant $G$ is then then determined by
	\begin{align}
		G &= \frac{1}{8\pi\phi_0^2} \quad ,
	\end{align}
	i.e. $\phi_0 = \kappa^{-1/2}$ recovers standard GR as coupled to the Standard Model.
	
	\subsection{Symplectic Structure and Weyl Algebra}
	
	For a graph $\Gamma$ with edges intersecting the boundary $S$ of a spatial section $\Sigma_0$ in  a causal double cone $\mathcal{D}$, we can define a Weyl algebra of observables localized thereon.
	
	A (spin network) state $\omega$ over the algebra ${\cal A}(\Sigma_0)$
	may be defined by representations on
	a closed, oriented differentiable {\em finite} graph $\gamma$
	embedded in $\Sigma_0$,
	with differentiable edges $e\in E$
	intersected transversally by a
	differentiable $2$-dimensional oriented surface $S$
	at a countable of intersection vertices $v\in V$.
	Let $C_{\Gamma}\in \Cyl$ be a $\Cinf$ Cylinder function with respect to the holonomy group $G$ on $\Gamma$, i.e. $C_{\Gamma}:=c(g_1,\ldots,g_N)$ where $g_k\in G$,
	and $c$ is a differentiable function.
	With test function $f$ a derivation
	$X_{S,f}$ on $\Cyl$ is defined by
	\beq{deriv}
	X_{S,f}\cdot C_{\Gamma}:=\frac{1}{2}
	\sum_{v\in V}\sum_{e_v\in E:\partl e_v\ni v}
	\kappa(e_v) f^i(v) X^i_{e_v} \cdot c ,
	\eeq
	where $\kappa(e_v)=\pm1$ above/below $S$
	(for the following purposes we may just exclude the tangential
	case $\kappa(e_v)=0$)
	and $ X^i_{e_v} \cdot c $ is the action of the left/right invariant
	vector field (i.e. $e_v$ is oriented away from/towards the
	surface $S$) on the argument of $c$ which corresponds to the
	edge $e_v$. Let $\Der$ denote the span of such derivations.
	
	Here the classical (extended) phase space is the cotangent
	bundle  $P:=T^*{\mathcal{C}}$
	over a space $\mathcal{A}$ of (suitably regular) finitely localized
	connections.
	Let $\delta = ( {\delta_A} , {\delta_E} ) \in T^*{\mathcal{C}}$.
	With suitable boundary conditions,
	a (weakly non-degenerate) symplectic form $\Omega$ over $T_e P$
	acts via
	\bear{sympl}
	\Omega|_{(A_e,E_e)}
	\left ( \delta, \delta' \right )
	&:=&
	\frac{1}{\ell^2} \int_{\Sigma} {\rm Tr}
	\left [  *E \wedge A'  -   *E' \wedge A  \right ] ,
	\ear
	with normalizing scale factor $\ell$. After lifting from $\mathcal{C}$ to phase space $P=T^*{\mathcal{C}}$,
	the cylinder functions $q\in\Cyl$ serve as
	(gauge invariant) configuration variables and
	the derivations $p\in\Der$ serve as conjugate momentum variables. They are obtained by the Poisson-Lie action
	of $2$-dimensionally smeared duals of densitized triads $E$.
	$\Cyl\times\Der$ has a Poisson-Lie structure
	\beq{Poisson}
	\{(q,p),(q',p'))\}:=(p q'- p' q  ,[p,p']) ,
	\eeq
	where $[p,p']$ denotes the Lie bracket of $p$ and $p'$.
	An antisymmetric bilinear form on $\Cyl\times\Der$ is given by
	\bear{pres}
	\Omega
	\left ( {(\delta_q,\delta_p)}, {(\delta_q',\delta_p')} \right )
	&:=&
	\int_{{\mathcal{C}_{\gamma\cup\gamma'}}/
		{\mathcal{G}_{\gamma\cup\gamma'}}}d\mu_{\gamma\cup\gamma'}
	\left [  p  q'  -   p' q  \right ] ,
	\ear
	where $q,q'\in \Cyl$ have support on $\Gamma$ resp. $\Gamma'$,
	with $p  q' -  p' q\in \Cyl$
	integrable over ${\mathcal{C}_{\Gamma\cup\Gamma'}}/
	{\mathcal{G}_{\Gamma\cup\Gamma'}}$ with
	measure $d\mu_{\Gamma\cup\Gamma'}$.
	
	With a symplectic form $\Omega$, 
	quantization of a function $\Omega(f,\cdot)$ yields a selfadjoint operator $\hat\Omega(f,\cdot)$
	and a corresponding
	unitary Weyl element $W(f):=e^{i\hat\Omega(f,\cdot)}$ can be constructed
	on a Hilbert space extension.
	With multiplication
	$W(f_1) W(f_2) := e^{i\Omega(f_1,f_2)}W(f_1+f_2)$,
	and conjugation $*: W(f)\mapsto W(-f)$,
	the Weyl elements generate a $*$-algebra.
	A norm on is defined by
	$\|f\|:=\frac{1}{4}\sup_{g\neq 0}\frac{\Omega(f,g)}{<g,g>}$.
	The $C^*$-closure under the operator sup-norm then generates
	a  $C^*$-algebra from the Weyl elements.
	
	In a gauge and $3$-diffeomorphism invariant
	representation of the algebra ${\cal A}_{\gamma}$ of  observables localized at $\gamma$,
	the observables in are represented by
	configuration multiplication operators $C_{\gamma}\in \Cyl$
	on Hilbert space $\mathcal{H}_{\gamma}$,
	and by gauge-invariant and 3-diffeomorphism invariant
	combinations with derivative operators $X_{S,f}\in \Der$.
	E.g., certain quadratic combinations yield
	the area operator with spectrum as in (32).
	
	Alternatively, using holonomy flux variables $(h_e[A], E_S^a[E] )$, for each edge $e$ of $\gamma$ intersecting the surface $S$ at puncture $p=e\cap S$, quantizations yield unitary operators
	\begin{align}
		U_e(\lambda) &= \hat{h}_e[\lambda A] \\
		V_e(\mu) &= e^{i\mu \hat{E}_e}  \quad ,
	\end{align}
	which do not commute at $p$ (cf. equations (28), (30), (31) above).
	Operator products from all intersection points $p$ then generate 
	an algebra $\mathcal{W}(\gamma, S)$ for the graph $\gamma\subset \Sigma_0$ on the boundary surface $S=\partial \Sigma_0$.

	\subsection{Quantum-like Evolution}
	
	In LQG, a solution of the Hamiltonian constraint \eqref{eq:hamiltonian_lqg} is required for covariant evolution lapses from one Cauchy surface to the next.
	However, for implementing a quantum like dynamics covariantly,
	various approaches exist , e.g.:
	\begin{itemize}
		\item The master constraint program \cite{thiemann2007}
		\item Spin foam models provide a covariant path integral formulation \cite{perez2013}
		\item Causal dynamical triangulations \cite{ambjorn2012}
	\end{itemize}
	
	\subsection{Relation to Causal Fermion Systems}
	
	As an alternative approach, causal fermion systems (CFS) start with fermions and derive spacetime causality from the fermionic structure \cite{finster2016,finster2020}. The key object is the fermionic projector
	\begin{equation}
		P(x,y) = \sum_n \psi_n(x) \otimes \overline{\psi_n(y)} ,
	\end{equation}
	where $\psi_n$ are fermionic wave functions.
	
	The causal structure emerges from the spectral properties of $P(x,y)$. With the correlation structure derived from $P(x,y)$, a connection between the respective spinor and tangent spaces
	of $x$ and $y$ is obtained. $x$ and $y$ can be projected to the spatial Cauchy section $\Sigma_0$.
	If $x$ and $y$ correspond to events, each with a fermion attached, then their $\Sigma_0$-projection onto a common graph yields vertices $v_x$ and  $v_y$, each with a spin $\frac{1}{2}$ edge,  $e_x$ resp. $e_y$, attached.   
	In CFS, like in the standard gauge theory approach, only the fermions are in the fundamental representation, while gauge bosons are acting in the adjoint representation between them.
	In a spin network representation, the spin on the edges can be both, odd or even. E.g. an intertwiner of edges of mixed spins, $\frac{1}{2}$ and $1$, is supposed to be a projection 
	of a fermion boson interaction. However, more detailed investigations are necessary to clarity the relations, between spin networks for geometry and the standard gauge theory of particles and the CFS approach.
	
	\section{Conclusions}
	
	We have reviewed approaches to spacetime geometry in terms of spinor variables.
	In any case, spinors are more fundamental than metric geometry:
	\begin{itemize}
		\item Already classically the metric, and more generally any tensor field describing the geometry, can be recovered in terms of spinor variables.
		\item All matter is spinorial (fermions inevitably).
		\item Spinors are inherent to the spin network approach of loop quantum gravity.
		Spin networks yield the holographic black hole entropy proportional to horizon area resulting from spinor intersections with the horizon sphere.
		\item Causal fermion systems recover (Minkowski) spacetime from a correlation structure from projection operators on fermions. 
	\end{itemize}
	
	The {\emph classical} spinor description of spacetime locally solders tangent vectors of the spacetime manifold to bispinorial tetrads. In order to cover the whole manifold $M$ globally,
	the latter has to admit a spin-structure in addition to and compatible with its differentiable topological structure.
	
	The {\emph 3+1 decomposition} of the spinor variables naturally yields spatial $SU(2)$-spinors.
	Their densification provide the dual variable to the orginal Ashtekar's self-dual connection.
	
	By integration along paths on a graph {\emph holonomy and flux variables} can be constructed. These generate a phase space of  UV-finite variables.
	
	{\emph Spin network states} solve the kinematic constraints and reveal a discrete geometry by the  eigenvalues of area and volume operators.
	
	A spin network may be contained in the spatial slice of a {\emph causal (double) cone}.  The size of the double cone is determined by a conformal scale factor. 
	For a conformal geometry, the geometric operators will be scale free on the spin network states.
	
	It is proposed that the non-geometric spin $0$ edges in the spin network arise as projections from Higgs scalars, which are supposed to provide mass to all particles in the interior of the causal double cone. Each massive particle path inside the double cone then is a Compton clock.
	The clock projected to the spin network provides a scale $l_c$ for the corresponding geometric operators, which may be larger than Plack scale $l_P$, but should be smaller than phase shifts detected under the influence of gravitational waves. Ultimately the average density of particle interaction vertices inside the spatial section $\Sigma_0$ and the density of particle spinors intersecting its boundary surface will set (in an appropriate average over the network) the regularization length for UV-regularization. In our approach, the regularization scale from the particle network (rather than the ad hoc scale $l_P$) enters the geometric operators on spin network states, and fixes their spectrum.  
	
	On the boundary of $\Sigma_0$ the Poisson algebra of holonomies and fluxes provide a {\emph net of local algebras on the graph} in the sense of algebraic quantum (field) theory.
	
	The sketchy developments in the current paper can be understood as proposal for further research along several open questions and into different possible directions.
	
	E.g. the Hamiltonian constraint in canonical LQG must be solved for full covariance, which still remains difficult. For that purpose the operator ordering ambiguity is usually fixed in a quite pragmatic way. 
	Experience from the Wheeler deWitt (WdW) equation suggests that, conformal covariance may solve the factor ordering for the scalar (also called Hamiltonian) constraint operator in a  unique way. E.g. on WdW minisuperspaces of geometry, the Laplace operator can be uniquely fixed to be the conformal one \cite{rainer1994conformal}.
	
	The conformal geometry and causal structure should be understood separately from the development of the volume measure of geometry. Massless particle modes per definition trace out geodesic paths. With increasing number of such path in a causal domain the causal cone structure of spacetime becomes more and more fixed.
	Smooth classical spacetime should be recovered in a limit of increasingly large spin networks. 
	
	Massive particle modes, on the other hand, provide natural Compton clocks defining the time and length scale of geometry, breaking the conformal invariance. 
	The Higgs mechanism on the particle interaction network should be understood in more detail, and in its projection to spin networks. Similarly the coupling of matter to classical geometry should be recovered from the particle spinor interaction network.
	
	The nature of the spin network and the geometrical operators thereon should be understood better: Is a spin network just a projection from a particle path network (in the Feynman sense), with quantum nature just inherited from the particle (matter) content? Or do they represent exclusive structures of space(time) geometry, as currently assumed by LQG?
	
	Clarifying the relation between spin networks (as in LQG) and correlation structures such as provided by fermionic projectors (as in causal CFS) might be helpful to clarify further, whether spacetime geometry can emerge quasi classically but discretely regularized, from a quasi classical content of an elementary particle spin network, or it has to be genuinely quantum with its own (spinorial or other) constituents. 
	
	We argue, that the former will be a much more natural and efficient choice: The spin networks should arise as entanglement networks of elementary particle interactions. The latter being ultimately quantum in nature, the induced spin networks will inherit the quantum nature from the particle content of the spacetime vacuum. That particular also explains the apparent similarity of classical Newtonian gravitational action at distance to quantum entanglement.
	
	The classical metric field can be locally captured by a bispinor-field.
	The quest for a topological, i.e. diffeomorphism invariant, spin-connection-based description of spacetime geometry has lead to spin networks on a graph with spin-representations on its edges and the representation intertwiners on its vertices. The discrete structure of geometry arises naturally, when diffeomorphism invariance and gauge invariance are combined in holonomy loops on discrete graphs. We have pointed out a novel approach to the spin networks underlying geometry. They may arise as a natural projection from the interaction structure of particle spinors, i.e. spinors belonging to the spin $0$, $\frac{1}{2}$ or $1$ representations of the SM. Note however that, the spin network in our current approach is different from a spatial network of entangled fermion particles at vertices and gauge bosons on connecting (tubular extensions of) edges as in \cite{rainer1994particlenet}. Also, in addition to the spin gauge group $SU(2)$, the gauge group of the SM generates additional spatial entanglements which have not been considered here. A more complete elaboration clarifying that relations is left to a forthcoming article.   
	
	Clarifying the relationships between different approaches such as loop quantum gravity, causal fermion systems, causal algebraic quantum (field) theory, topological and conformal QFT, and string theory concepts may help to forge a unified understanding.
	
	Spacetime itself is a very successful concept, but emergent only at scales above the Planck length. From an atomistic perspective, its geometry must be derived from fundamental microscopic constituents. 
	We have proposed that, a causal interaction network of all the elementary spinor particles ever created or annihilated in a defined causal domain may, together with their interactions, naturally generate the spin network which then provides their underlying UV-regularized space(time) geometry.
	
	\section*{Acknowledgments}
	
	I thank all colleagues from the 'Causal Fermion Systems 2025' conference at Regensburg University for inspiring discussions, in particular to Claudio Paganini for some useful comments. This paper is a follow up of a presentation at the 'Modern Trends in Physics 2025' conference at Baku State University on 27.11.2025.

\end{document}